\documentclass[prb,twocolumn,showpacs,amsmath,amssymb,superscriptaddress]{revtex4}
\usepackage[pdftex]{graphicx}
\usepackage[outdir=./]{epstopdf}
\usepackage{enumerate,ulem}
\newcommand \beq{\begin{eqnarray}}
\newcommand \eeq{\end{eqnarray}}

\newcommand{\bm}[1]{\boldsymbol{#1}}

\newcommand{\bfr}{\bm{r}}
\newcommand{\bfk}{\bm{k}}

\newcommand{\bfM}{\bm{M}}
\newcommand{\bfp}{\bm{p}}
\newcommand{\bfsigma}{\bm{\sigma}}

\newcommand{\bmc}[2]{\left( \begin{array}{c} #1  \\ #2 \end{array}\right)}

\newcommand{\sech}{\mathrm{sech}}

\begin{document}

\title{Universal charge and current on magnetic domain walls in Weyl semimetals}
\author{Yasufumi Araki}
\affiliation{Institute for Materials Research, Tohoku University, Sendai 980-8577, Japan}
\affiliation{Frontier Research Institute for Interdisciplinary Sciences, Tohoku University, Sendai 980-8578, Japan}
\author{Akihide Yoshida}
\author{Kentaro Nomura}
\affiliation{Institute for Materials Research, Tohoku University, Sendai 980-8577, Japan}

\begin{abstract}
Domain walls in three-dimensional Weyl semimetals, formed by localized magnetic moments, are investigated.
There appear bound states around the domain wall with the discrete spectrum,
among which we find ``Fermi arc'' states with the linear dispersion.
The Fermi arc modes contribute to the electric charge and current localized at the domain wall,
{which reveal} a universal behavior depending only on chemical potential and the splitting of the Weyl nodes.
{This equilibrium current can be traced back to the chiral magnetic effect, or the edge counterpart of the anomalous Hall effect in the bulk.}
We propose a new way to manipulate the motion of the domain wall, accompanied with the localized charge,
by applying an external electric field.
\end{abstract}

\maketitle

\section{Introduction}
Domain wall (DW) in ferromagnetic materials is one of the key ingredients in spintronics \cite{Tatara},
which can be made use of as a carrier of information.
Racetrack memory technology, experimentally realized in 2008,
uses magnetic domains in ferromagnetic nanowires for high-performance information storage \cite{Parkin}.
It is essential for such applications to control the motion of DWs;
besides the application of external magnetic fields,
spin-transfer torque coming from a spin-polarized current via the $s$-$d$ exchange interaction
is one of the major approaches to drive DW motion \cite{Berger,Saihi,Slonczewski}.
An alternative way has recently been proposed by using spin-orbit torques,
which are induced in spin-orbit-coupled materials by the Rashba--Edelstein effect and the spin Hall effect \cite{Manchon_SOT,Obata,Matos-Abiague,Gambardella,Fert,Ryu,Emori,Tserkovnyak_2012,Shiomi_2014}.

It has been widely expected that topological materials,
such as topological insulators (TIs), Dirac semimetals (DSMs), Weyl semimetals (WSMs),
can play important roles in spintronic applications \cite{Pesin_MacDonald}.
They are characterized by the band-touching ``Dirac cone'' structure,
which is realized by strong spin-orbit coupling.
TIs exhibit two-dimensional (2D) Dirac cone structure on their surfaces \cite{Hasan_Kane,Qi_Zhang},
while DSMs and WSMs have 3D Dirac cone structure in the bulk,
with or without degeneracy required by time-reversal and inversion symmetries \cite{Young_2012,Wan_2011,Burkov_2011}.

The band topology along with magnetism gives rise to various kinds of anomalous magnetoelectric properties,
such as the anomalous Hall effect (AHE)  \cite{Qi_2008,Yu_2010,Nomura_2011,Checkelsky_2012,Chang_2013,Burkov_2014,Burkov_2014_2}, {the chiral magnetic effect (CME)} \cite{Vilenkin,Fukushima_2008,Kharzeev_2008,Kharzeev_2014}, etc.
Magnetic TIs and WSMs show quantized and non-quantized AHE, respectively,
due to breaking of time-reversal symmetry.
It should be noted that the topological nature is imprinted in their boundary as well as the bulk,
arising as the dissipationless chiral edge current \cite{Wang_2013},
like the quantum Hall states under magnetic fields \cite{Laughlin,Halperin}.
Hence we can expect that the magnetic domain boundaries in those topological materials can bear
properties richer than those in normal magnets,
either static or dynamic.
TIs coupled to ferromagnetic DWs have been studied in recent {literature},
exhibiting a 1D chiral channel at the DW that contributes to anomalous magnetoelectric transport and DW dynamics \cite{Tserkovnyak_2012,Nomura_2010,Wickles_2012,Ferreiros_2015,Wakatsuki_2015,Tserkovnyak_2015,Upadhyaya_2016}.
Magnetic DWs in WSMs might be even more complex, due to their higher dimensionality.

In this paper, we study the properties of magnetic DWs,
formed by localized magnetic moments in 3D WSMs,
by exactly solving the Weyl equation under the DW texture.
We find a number of localized modes at the DW,
even though there is no mass gap in the bulk.
The spectrum of the bound states is discretized,
as a result of the Landau quantization
under an ``axial'' magnetic field generated by the magnetic texture.
The topological characteristics are obviously imprinted in the zeroth Landau level (LL) among them:
its band crosses zero energy by an open line in the momentum space,
called ``Fermi arc'',
which connects two Weyl points in the 2D Brillouin zone projected on the DW,
like the surface states of WSMs \cite{Wan_2011}.
The Fermi arc states give rise to the universal charge and current localized at the DW,
which can be regarded as {the CME under the axial magnetic field},
 or the edge counterpart of the AHE in the bulk WSM \cite{Burkov_2014_2}.
Such a localized charge can enable us to manipulate the motion of the DW,
the velocity of which can be tuned via the chemical potential.
In the following calculation, we take $\hbar=1$ and restore it in the numerical results.

\begin{figure}[tbp]
\includegraphics[width=8cm]{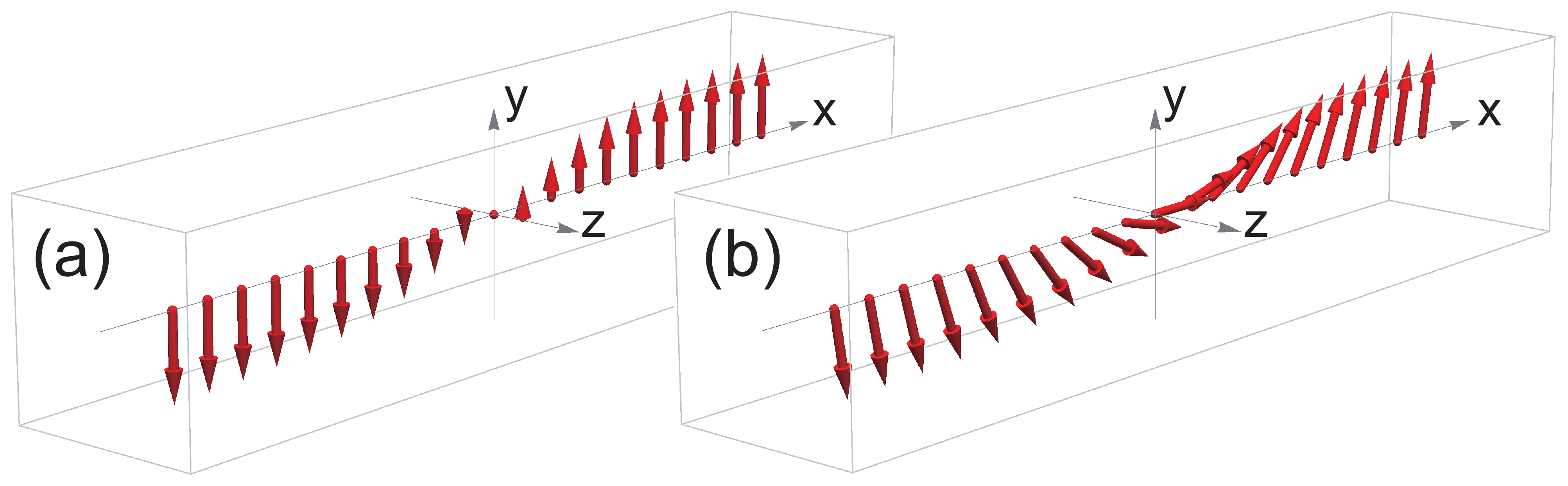}
\caption{Schematic pictures of the DW configurations treated in this paper: (a) the simple collinear DW and (b) the Bloch DW.}
\label{fig:domainwall}
\end{figure}

\section{Model}
We start with the minimal continuum Hamiltonian for the electrons in WSMs \cite{Kurebayashi_2014},
\begin{align}
\hat{H}(\bfr) = v_F \tau_z \bfsigma \cdot \hat{\bfp} - J \bfM(\bfr)\cdot\bfsigma,
\end{align}
where $v_F$ is the Fermi velocity and $\hat{\bfp} \equiv -i \bm{\nabla}$ is the momentum operator.
The Hamiltonian acts on the 4-component spinor $\psi(\bfr)$ in the spin and pseudospin (chirality) spaces,
with the Pauli matrices $\bfsigma$ and $\bm{\tau}$, respectively.
The continuous vector field $\bfM(\bfr)$ represents localized magnetic moments,
coupled to the Weyl electrons via the exchange interaction strength $J$.
{It is proposed that such a magnetic Weyl semimetal can be realized
 by some cobalt-based Heusler and half-Heusler alloys} \cite{Kubler_2016,Bernevig_2016,Hasan_2016}.
The local magnetization here can be regarded as an ``axial gauge field'' $\bm{a}(\bfr) = (J/v_F)\bfM(\bfr)$,
which serves as a gauge potential with opposite signs for opposite chiralities,
defined by $\hat{H} = v_F \tau_z \bfsigma \cdot \left[\hat{\bfp} - \tau_z \bm{a}(\bfr) \right]$.

\begin{figure}[tbp]
\includegraphics[width=8.5cm]{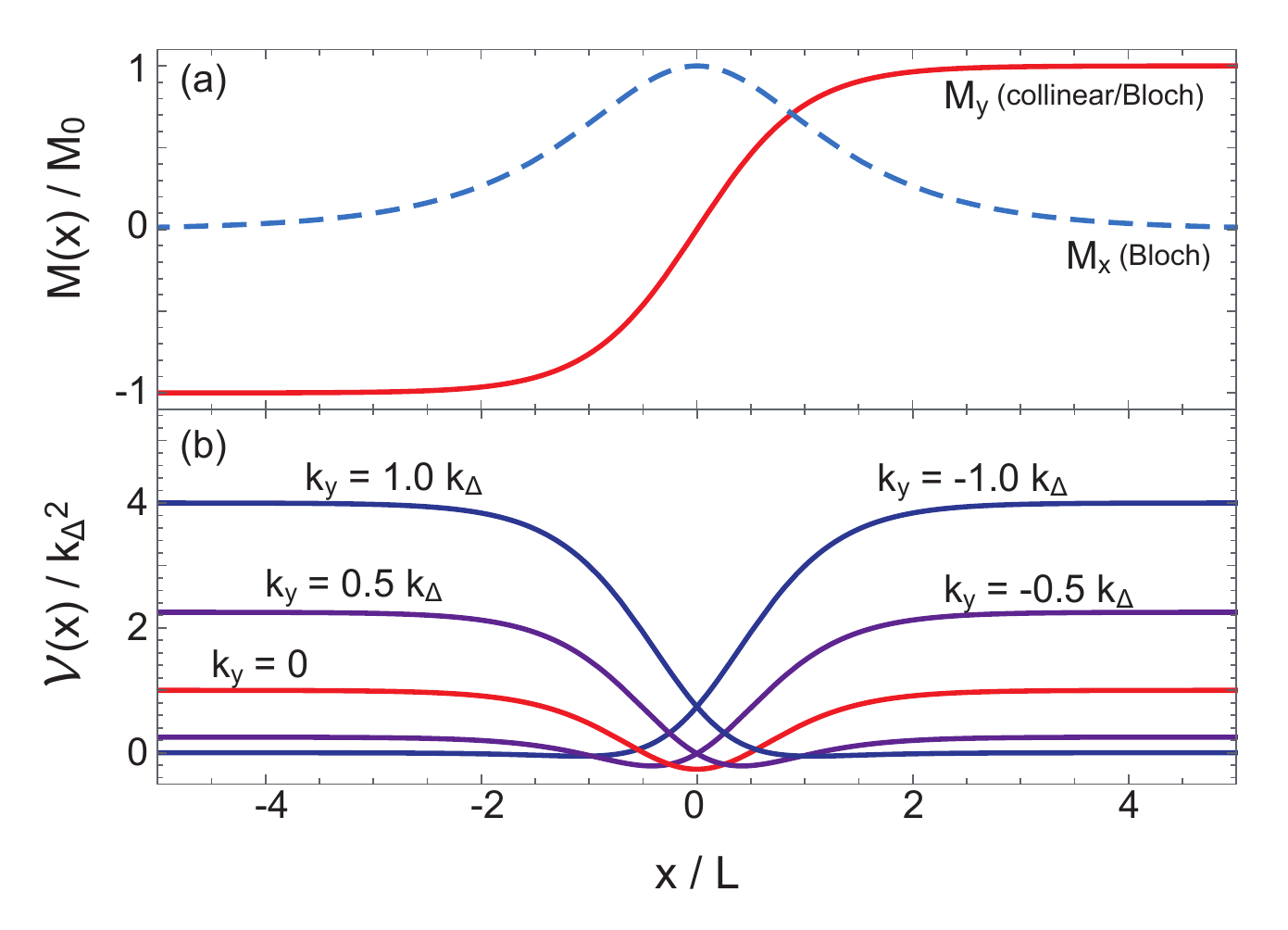}
\caption{Spatial profiles of (a) the magnetization $\bfM(x)$ and (b) the ``potential'' $\mathcal{V}(x)$.}
\label{fig:current}
\end{figure}

Here we introduce the magnetic DW texture in $\bfM(\bfr)$, located at $x=0$ with the width $L$.
We take here the simple collinear DW configuration
\begin{align}
\bfM_\mathrm{coll}(\bfr) = M_0 \left( 0,\tanh\frac{x}{L},0 \right)
\end{align}
as shown in Figs.~\ref{fig:domainwall}(a) and \ref{fig:current}(a),
in which the magnetization is paramagnetic at $x=0$.
The Bloch-type DW configuration shown in Fig.~\ref{fig:domainwall}(b),
\begin{align}
\bfM_\mathrm{Bloch}(\bfr) = M_0\left(\sech\frac{x}{L},\tanh\frac{x}{L},0\right),
\end{align}
reduces to $\bfM_\mathrm{coll}(\bfr)$ by local U(1) chiral gauge transformation
$\psi(\bfr) \mapsto e^{i\tau_z\phi(x)}\psi(\bfr)$,
with the phase factor $\phi(x) = (JM_0/v_F) \sech (x/L)$.
Both of these configurations drop to an asymptotically uniform magnetization $M_y \sim \pm M_0$ in the bulk far away from the DW,
which shifts the Weyl points to $k_y = \pm k_\Delta$, with $k_\Delta \equiv JM_0/v_F$.
Such DW textures give rise to the ``axial magnetic field'' corresponding to the rotation of the axial gauge field,
\begin{align}
\bm{b}(\bfr) = \bm{\nabla} \times \bm{a}(\bfr) = \left(0,0,\frac{k_\Delta}{L}\sech^2\frac{x}{L}\right),
\end{align}
along the DW.

We search for the eigenstates and the spectrum under the DW configuration $\bfM_\mathrm{coll}(\bfr)$,
by solving the Weyl equation $\hat{H}(\bfr) \psi(\bfr) = E \psi(\bfr)$.
Due to translational symmetry in $y$- and $z$- directions,
we can use the plane-wave basis with the wave vector $\bfk_\perp = (k_y,k_z)$.
Since the left and right chiralities (Weyl nodes) are decoupled,
what we need to solve is the 2-component characteristic equation in one dimension,
\begin{align}
\left[-i \tau_z\sigma_x \partial_x +(\tau_z k_y -k_\Delta \xi)\sigma_y + \tau_z k_z \sigma_z -\epsilon \right] \psi^{(\tau_z)} = 0, \label{eq:Weyl-DW}
\end{align}
where $\epsilon \equiv E/v_F$ and $\xi \equiv \tanh(x/L)$.
The 2-component spinor $\psi^{(\tau_z)}=(u^{(\tau_z)},v^{(\tau_z)})^T$ spans the spin SU(2) subspace,
with the quantization axis taken in $s_z$-direction.
Here we first solve the equation for the chirality $\tau_z = +1$ with the superscript $(\tau_z)$ suppressed,
and supplement the results for the opposite chirality later on.
All the numerical results in the figures are calculated with the parameters fixed to $L=100\mathrm{nm}, v_F=10^{6}\mathrm{m/s}$, and $J M_0=25\mathrm{meV}$.
Thus $k_\Delta = 0.038 \mathrm{nm}^{-1}$, and the maximum strength of $\boldsymbol{b}$ is $0.25 \mathrm{T}$.

\begin{figure}[tbp]
\includegraphics[width=9cm]{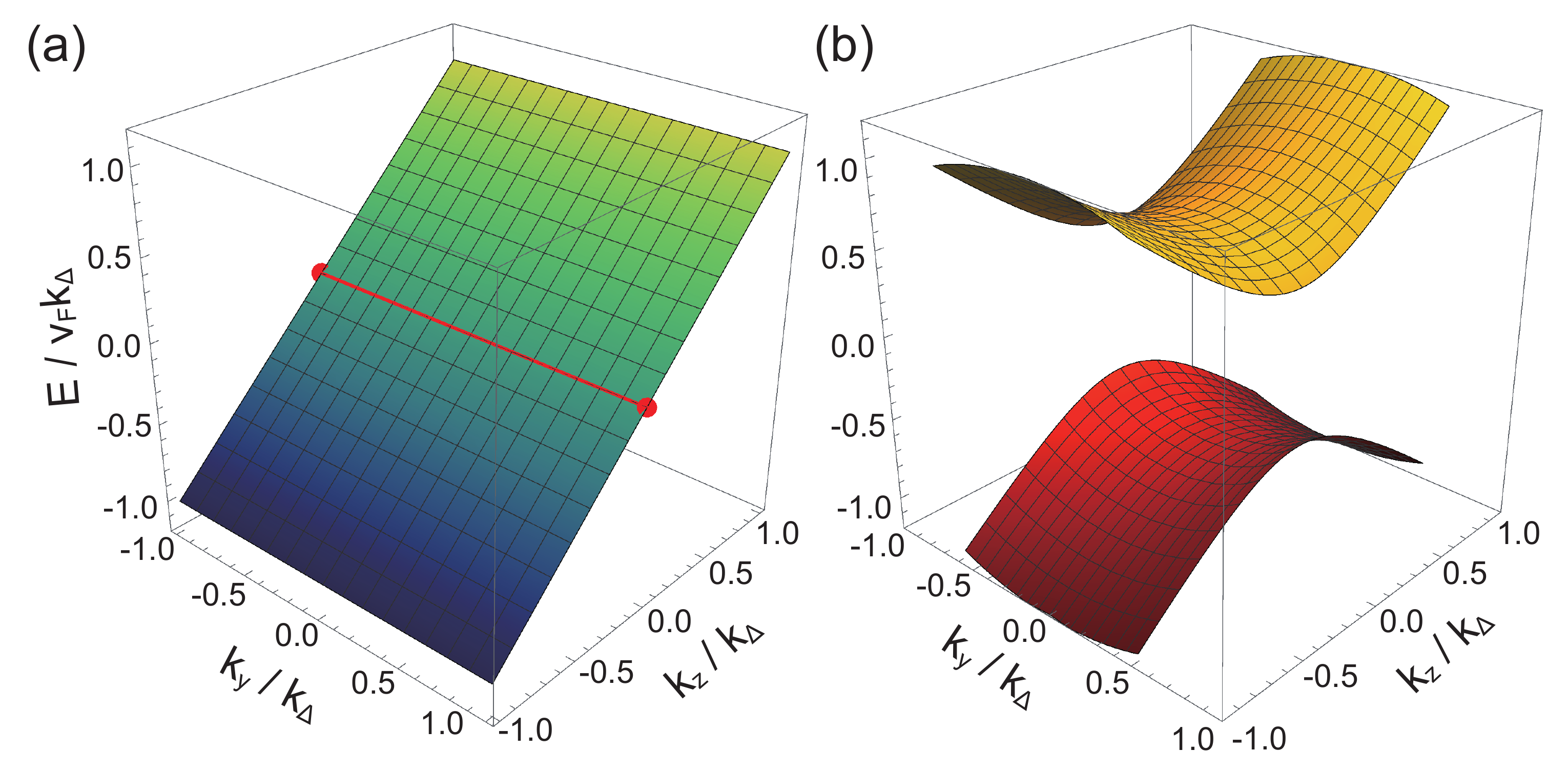}
\caption{The band structures of the bound states at the DW,
with the band indices (a) $N=0$ (``Fermi arc'' state) and (b) $N=1$.
The red line and points in (a) denote the Fermi arc and the Weyl points, respectively.}
\label{fig:bands}
\end{figure}

\section{Bound states and spectrum}
In normal WSMs, the electronic states cannot localize under scalar potentials,
which is known as Klein tunneling.
Here we search for bound states in the presence of the DW,
based on Eq.~(\ref{eq:Weyl-DW}).
This equation reduces to the non-relativistic Schr\"{o}dinger-like equation for a single component \cite{Cooper_SUSY},
\begin{align}
\left[-\partial_x^2 + \mathcal{V}(x) \right]u(x) = \mathcal{E} u(x), \label{eq:Weyl-Schrodinger}
\end{align}
where
\begin{align}
\mathcal{V}(x) \equiv -\frac{k_\Delta}{L} \sech^2\frac{x}{L} + \left(k_y-k_\Delta\tanh\frac{x}{L}\right)^2
\end{align}
and $\mathcal{E} \equiv \epsilon^2 - k_z^2$.
As long as $|k_y| < k_\Delta + L^{-1}$,
the ``potential'' $\mathcal{V}(x)$ has a pocket around the DW that satisfies $\partial_x \mathcal{V}(x)=0$,
as shown in Fig.~\ref{fig:current}(b).
This implies that there exist bound states in the pocket (around the DW),
even though there is no mass gap in the bulk.
In the limits $x \rightarrow \pm\infty$, the potential becomes asymptotically flat,
i.e. $\mathcal{V}(x) \sim (k_y \mp k_\Delta)^2$.
Thus the asymptotic behavior of the bound states can be easily obtained,
showing the exponential decay behavior $u(x),v(x) \sim \exp(\mp \kappa^\pm x)$,
where the decay rate $\kappa^\pm =\left[ k_z^2 + (k_y \mp k_\Delta)^2 -\epsilon^2 \right]^{1/2}$.

\begin{figure}[tbp]
\includegraphics[width=8.5cm]{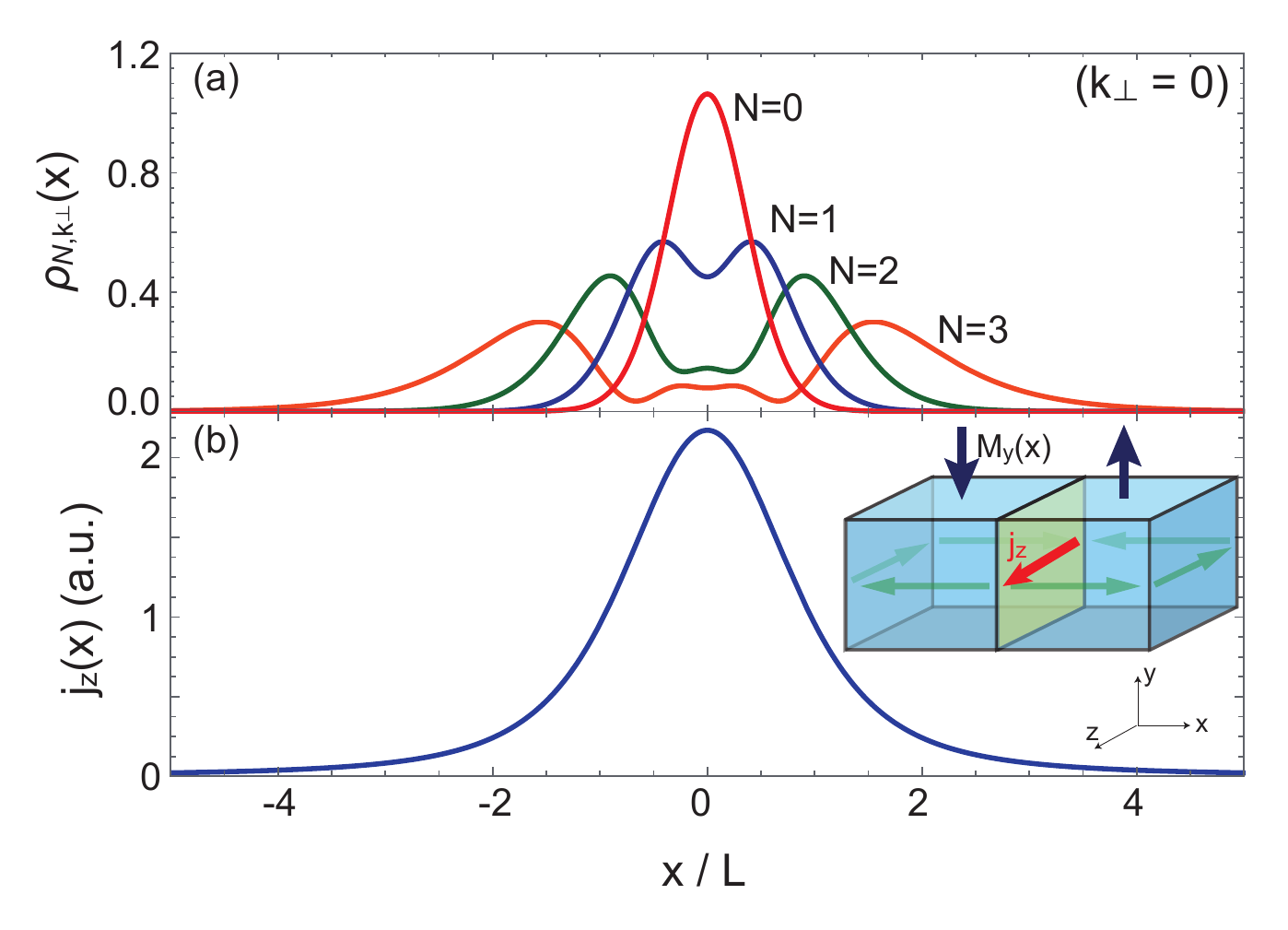}
\caption{Spatial profiles of (a) the probability distribution $\rho_{N,\bfk_\perp}(x)$ with $\bfk_\perp=0$ and (b) the equilibrium current $j_z(x)$.
Here the upper limit of $N$, given by $\lfloor k_\Delta L \rfloor$, is 3.
The inset shows the schematic picture of the equilibrium current at the DW.}
\label{fig:wf-distribution}
\end{figure}

Equation (\ref{eq:Weyl-Schrodinger}) reduces to an already-known hypergeometric differential equation by change of variables,
which can be analytically and exactly solved (see the Appendix for detailed calculations).
Among the exact solutions,
we find a number of bound states with the exponential decay behavior,
showing the discrete spectrum labeled by an integer $N \in [0,k_\Delta L)$.
There are one linearly dispersed mode with $N=0$ (see Fig.~\ref{fig:bands}(a)),
\begin{align}
\epsilon_0(\bfk_\perp) = k_z,
\end{align}
and several other modes labeled by nonzero $N$ (see Fig.~\ref{fig:bands}(b) for $N=1$),
\begin{align}
\epsilon_N(\bfk_\perp) = \pm \left[ k_z^2 + (k_\Delta^2/\bar{\kappa}_N^2 -1)(\bar{\kappa}_N^2 - k_y^2) \right]^{1/2},
\end{align}
where $\bar{\kappa}_N = k_\Delta - N/L$.
The chirality $\tau_z=-1$ gives the same spectrum, leading to the twofold degeneracy.
The wavefunctions $\psi_{N,\bfk_\perp}(x)$ for these eigenvalues can be obtained exactly
in terms of hypergeometric functions (see the Appendix for their exact forms).
The dimensionless probability distribution $\rho_{N,\bfk_\perp}(x) \equiv L |\psi_{N,\bfk_\perp}(x)|^2$, with $\bfk_\perp$ fixed to zero,
is shown in Fig.~\ref{fig:wf-distribution}(a) for each $N$;
it shows $N+1$ peaks,
which is consistent with the potential pocket picture in Eq.~(\ref{eq:Weyl-Schrodinger}).

Since the wavefunctions for the bound states should exponentially decay away from the DW
(i.e. $\kappa^\pm > 0$),
$k_y$ is limited in the region $|k_y| < \bar{\kappa}_N^2/k_\Delta$ for each $N$;
otherwise the wavefunction shows the oscillatory behavior, which corresponds to the extended state
(see Eqs.~(\ref{eq:app-k1})-(\ref{eq:app-k-cond}) in the Appendix).
This condition is stricter than that qualitatively obtained from the potential pocket picture ($|k_y| < k_\Delta + L^{-1}$).
The $N \neq 0$ modes have saddle points at $\bfk_\perp =0$,
which gives rise to van Hove singularity in the density of states.

\begin{figure}[tbp]
\begin{tabular}{ll}
\includegraphics[width=4.0cm]{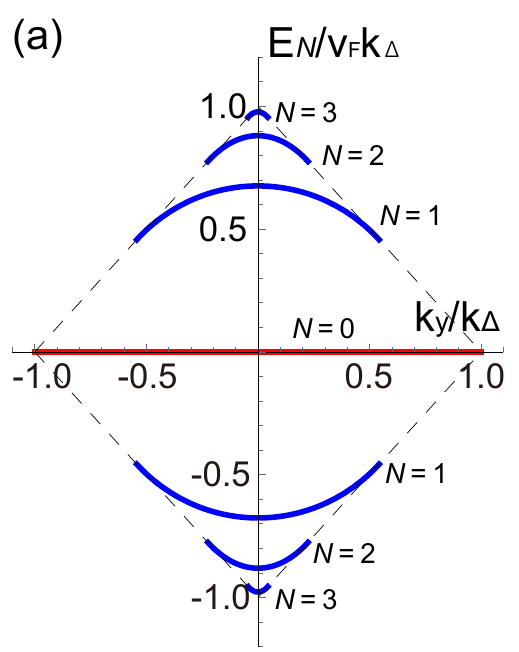} & 
\includegraphics[width=4.1cm]{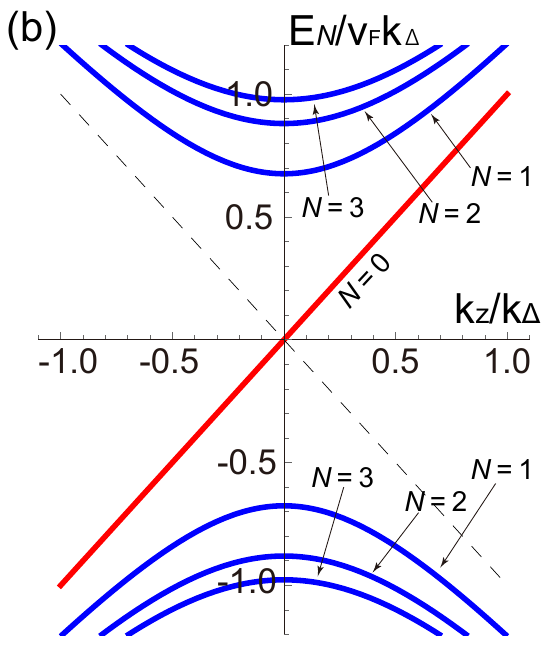}
\end{tabular}
\caption{The spectrum of the bound state at the DW, with one of the momentum components,
(a) $k_z$ or (b) $k_y$, fixed to zero.
The red line corresponds to the $N=0$ mode, i.e. the ``Fermi arc'' state.
}
\label{fig:spectrum}
\end{figure}

The discretized spetrum of the DW-bound states in WSMs
has a structure quite similar to that found in the surface states of TIs with magnetic DWs \cite{Wickles_2012}:
the dispersion in $k_y$-direction, shown in Fig.~\ref{fig:spectrum}(a), is convex toward $|E|\rightarrow\infty$ like that in TIs with in-plane magnetization,
while the dispersion in $k_z$-direction, Fig.~\ref{fig:spectrum}(b), is concave like that in TIs with out-of-plane magnetization.
That is because the 3D Weyl Hamiltonian reduces to the 2D Hamiltonian of TI surface states by fixing one of the momentum components to zero.
In other words, the bound-state spectrum found in WSMs can be regarded as a 3D hybrid of two characteristic spectra known in TI surfaces.

The discretization of the spectrum shown here can be regarded as Landau quantization
under the axial magnetic field $\bm{b}(\bfr)$ applied in $z$-direction.
It should be contrasted with Landau quantization in WSMs under normal magnetic fields;
the zeroth LLs $(N=0)$ under normal magnetic fields are linearly dispersed along the magnetic field,
in antiparallel directions for opposite chiralities.
The axial magnetic field, on the other hand, couples to each chirality antiparallelly,
which makes their dispersion parallel \cite{Liu_2013}.
The number of LLs is limited here,
because the axial magnetic field is present only in the limited area around the DW.
Higher LLs penetrate into the continuum as the DW becomes thinner,
while the zeroth LLs remain stable {independently of $L$}.

\section{Fermi arc states}
Let us investigate the properties of the zeroth LLs in more details,
since they strongly reflect the topological nature of the system.
The linearly dispersed band crosses zero energy by an open line called ``Fermi arc'',
which connects the two Weyl points projected on the DW,
 $(k_y,k_z) = (\pm k_\Delta,0)$.
Such a Fermi arc structure is similar to that seen on the surfaces of WSMs \cite{Wan_2011},
while here it is twofold degenerate.

\begin{figure}[tbp]
\includegraphics[width=8.5cm]{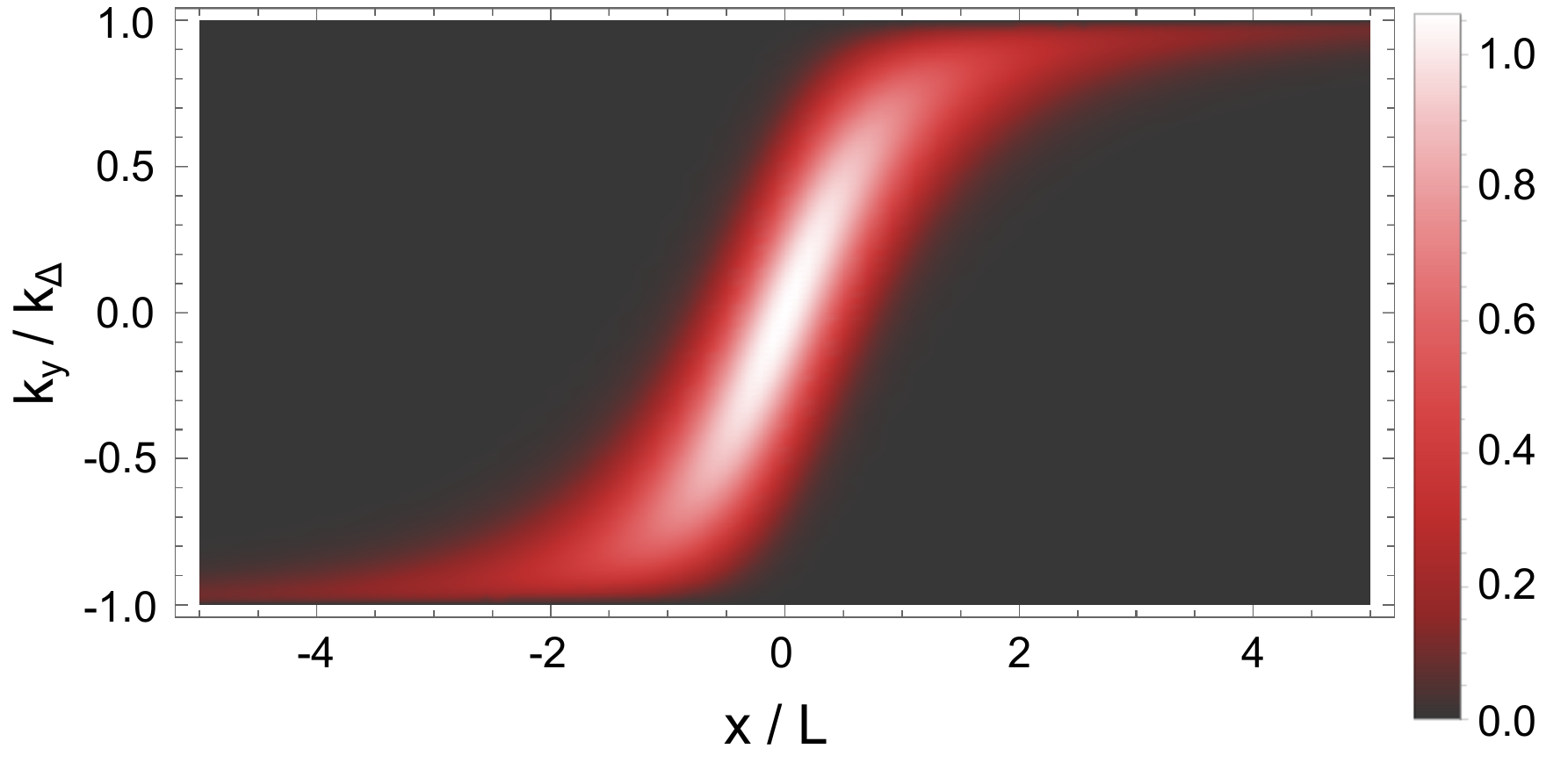}
\caption{The probability distribution $\rho_{0,\bfk_\perp}(x)$ of the DW Fermi arc states with the chirality $\tau_z =+1$.
It should be noted that $\rho_{0,\bfk_\perp}(x)$ is independent of $k_z$.}
\label{fig:fermiarc-coll-wf}
\end{figure}

The spatial profile of the DW Fermi arc state is given by the wavefunction
\begin{align}
\psi_{0,\bfk_\perp}^{(\tau_z)}(x) = e^{\tau_z k_y x} \left(\cosh \frac{x}{L}\right)^{-k_\Delta L} \chi^{(\tau_z)}, \label{eq:wf-0}
\end{align}
which is independent of $k_z$ but strongly depends on $k_y$.
Here $\chi^{(+)} = (1,0)^T$ and $\chi^{(-)} = (0,1)^T$.
Hence, the electron spin on the Fermi arc with the chirality $\tau_z=+1$ is fully polarized in $+z$-direction,
while the opposite polarization for the opposite chirality.
The dimensionless probability distribution $\rho_{0,\bfk_\perp}^{(\tau_z)}(x)$ for the chirality $\tau_z =+1$
is shown in Fig.~\ref{fig:fermiarc-coll-wf};
the peak for $k_y \sim 0$ is located in the vicinity of the DW,
while the wavefunction in the ends of the Fermi arc, i.e. $k_y \rightarrow \pm k_\Delta$,
extends in the bulk of half spaces $(x\rightarrow\pm\infty)$.
{This means that zeroth LLs are distributed
along the trajectory of the Weyl points, from $(k_y,k_z)=(-\tau_z k_\Delta,0)$ in $x\rightarrow -\infty$
to $(+\tau_z k_\Delta,0)$ in $x\rightarrow +\infty$.
The Fermi arc structure found here is the projection of the trajectory onto the $(k_y,k_z)$-plane.}
Such kind of Fermi arc is also seen at the interface of superfluid $^3$He-A phases \cite{Silaev_2012}.

\section{Localized charge and current}
Since the Fermi arc states at the DW, corresponding to the zeroth LL,
 are linearly dispersed in $+k_z$-direction {}{with the velocity $v_F$} for both chiralities,
they are robust under charge disorder and can contribute to equilibrium current,
while the other LLs do not contribute due to the dispersion symmetric around $\bfk_\perp=0$.
The equilibrium current density in $z$-direction at the position $x$ comes from all the occupied states in the zeroth LLs below the chemical potential $\mu$,
\begin{align}
j_z(x) &= e v_F \sum_{\tau_z} \sum_{\begin{subarray}{c} k_y \in (-k_\Delta,k_\Delta) \\ k_z \in (-k_C,k_F) \end{subarray}} \left| \psi_{0,\bfk_\perp}^{(\tau_z)}(x)\right|^2.
\end{align}
$k_F \equiv \mu/v_F$ is the Fermi wavenumber,
and $k_C (\equiv \Lambda/v_F)$ is the momentum cutoff in $k_z$-direction,
which is related to the lattice constant $a$ by $k_C \sim 1/a$.
Since the probability density $\rho_{0,\bfk_\perp}^{(\tau_z)}(x)$ is independent of $k_z$,
we obtain
\begin{align}
j_z(x) = \frac{e v_F}{(2\pi)^2}\frac{k_F + k_C}{L} \int_{-k_\Delta}^{k_\Delta} dk_y \sum_{\tau_z} \rho_{0,\bfk_\perp}^{(\tau_z)}(x), \label{eq:DW-current}
\end{align}
which is straightforwardly related to the local charge density $q(x)$ by $j_z(x) = e v_F q(x) + \mathrm{const.}$
The constant term here is determined so that the total charge $Q= \int dx\; q(x)$ for $\mu=0$ should be zero,
from particle-hole symmetry.
Thus the total charge (per unit area) and current (per unit length) localized at the DW are given by
\begin{align}
Q = \frac{e}{\pi^2} \frac{k_\Delta}{v_F} \mu, \quad I_z = \frac{e}{\pi^2} k_\Delta (\mu+\Lambda),
\label{eq:total-current}
\end{align}
respectively.
$Q$ and $I_z$ linearly depend on the chemical potential $\mu$
and the Weyl-point splitting $k_\Delta$,
with the universal coefficient $e/\pi^2$,
{while they are independent of the DW width $L$}.


The current density $j_z(x)$ has a sharp peak at $x=0$
as shown in Fig.~\ref{fig:wf-distribution}(b),
which means that the equilibrium current flows mainly on the DW in the direction perpendicular to the magnetization.
{Since the origin of this current is the zeroth LLs,
it can be reinterpreted as the anomaly-induced magnetoelectric response in WSM.}
{In the context of the usual CME} \cite{Fukushima_2008,Kharzeev_2008},
{a normal magnetic field $\boldsymbol{B}$ in the presence of the chiral chemical potential $\mu_5$,
namely the chemical potential difference between two chiralities,
induces an equilibrium current $\boldsymbol{j} = (e^2/2\pi^2)\mu_5 \boldsymbol{B}$.}
{Here the chemical potential $\mu$ is normal and the magnetic field $\boldsymbol{b}$ is chiral,
leading to the similar relation}
\begin{align}
\boldsymbol{j} =\frac{e^2}{2\pi^2}(\mu +\Lambda) \boldsymbol{b},
\end{align}
{where the constant shift $\Lambda$ to the chemical potential means
that the equilibrium current comes from all the occupied states down to the band bottom.
Integrating this relation over the region $-\infty < x < \infty$,
we obtain $I_z$ in Eq.~(\ref{eq:DW-current}).}

{Alternatively, it can be intuitively understood by following the idea of the AHE.}
Regarding the system as a junction of two WSMs magnetized in opposite directions,
each part shows the AHE
with the conductivity $\sigma_{xz}(x \gtrless 0)= \pm  (e^2/2\pi^2)k_\Delta$,
 due to the breaking of time-reversal symmetry \cite{Burkov_2014_2,Kurebayashi_2014}.
It accompanies the equilibrium current on the side surfaces,
circularly flowing in the direction perpendicular to the magnetization.
By attaching two parts side by side,
there remains a current flowing at the interface,
as shown in the inset of Fig.~\ref{fig:wf-distribution},
which is captured as a DW equilibrium current in the above calculation.

The relation between the DW equilibrium current and the AHE can be understood in terms of the Streda formula \cite{Streda},
\begin{align}
\sigma_{xz} = e \frac{\partial m_\mathrm{orbit}^y}{\partial \mu}. \label{eq:streda}
\end{align}
The orbital magnetization $\bm{m}_\mathrm{orbit}$ is related to the current density by
\begin{align}
\bm{j} = \bm{\nabla} \times \bm{m}_\mathrm{orbit},
\end{align}
and the local anomalous Hall conductivity, defined at long wavelength, arises from breaking of time-reversal symmetry by
\begin{align}
\sigma_{xz}(x) = \frac{e^2}{2\pi^2} a_y(x).
\end{align}
Using these three relations, we obtain
\begin{align}
\frac{\partial j_z }{\partial \mu} = \frac{\partial (\partial_x m_\mathrm{orbit}^y)}{\partial \mu} = \frac{1}{e} \partial_x \sigma_{xz} = \frac{e}{2\pi^2} \partial_x a_y, \label{eq:anomalous-hall}
\end{align}
which qualitatively accounts for the peak structure at $x=0$ and the linear dependence on $\mu$ (or $k_F$),
shown by Eq.~(\ref{eq:DW-current}).
Integrating over $-\infty<x<\infty$, we obtain the relation for the total current $I_z$,
\begin{align}
\frac{\partial I_z}{\partial \mu} = \frac{e}{2\pi^2} a_y \bigr{|}_{x=-\infty}^{x=\infty} = \frac{e}{\pi^2} k_\Delta,
\end{align}
which agrees with Eq.~(\ref{eq:total-current}).
Thus we can understand that the equilibrium current along the DW can be traced back to
{the CME or} the AHE in the bulk WSM.

The existence of the localized charge $Q$ means that the DW is sensitive to the application
of an external electric field $E$.
{Since the Fermi surfaces in the left and right sides of the DW are well separated under a sufficiently large magnetization,
the electrons cannot be transmitted through the DW except for the Fermi arc states,
which implies that the conduction electrons cannot contribute to any torques on the DW magnetization,
such as the spin transfer torque or the spin-orbit torque.
Hence, the electrostatic force $QE$ on the DW is the only driving force for the DW motion
that can be generated by an electric field,
without any dissipation.}

{In the absence of a pinning potential,
the DW obeys the classical equation of motion} \cite{Doring},
\begin{align}
M_W \ddot{X} +\frac{M_W}{\tau_W} \dot{X} = Q E,
\end{align}
{where $X$ is the collective coordinate of the DW in $x$-direction.
The DW effective mass $M_W$ and the relaxation time $\tau_W$ are given by
$M_W = 2 / a^3 L K$ and $\tau_W = \alpha K$, respectively,
where $\alpha$ is the Gilbert damping constant and $K$ is the magnetic anisotropy energy} \cite{Tatara}.
{By applying an electric field $E$ in $x$-direction,
the DW eventually reaches the drift velocity}
\begin{align}
V_D = QE\frac{\tau_W}{M_W} = \frac{a^3 L k_\Delta k_F}{2\pi^2 \alpha}eE,
\end{align}
{as long as the DW motion is adiabatic so that there is no macroscopic change in the DW structure.
Taking typical values $\mu=20\mathrm{meV}, \ a=1\mathrm{nm}, \ \alpha = 0.01$ and $E=100 \mathrm{V/cm}$,
we can estimate the drift velocity $V_D \sim 10 \mathrm{m/s}$.}

{Since the system here is not in the insulating regime,
the localized charge $Q$ is subject to screening and the above phenomena can be suppressed to some extent.
However,
as the density of states $\rho \sim \mu^2$ in 3D WSMs,
in contrast to $\sim \mu^{1/2}$ in normal metals and semiconductors,
the Thomas-Fermi screening length is relatively long at low energy.
Therefore, the screening effect on the localized charge here is sufficiently small
in nanoscale systems that we are interested in.}

\section{Conclusion}
In this paper, we have seen the properties of magnetic DWs in WSMs.
There are a number of bound states around the DW,
among which we have found twofold degenerate linearly dispersed bands showing the Fermi arc structure.
They give rise to a universal charge and current localized at the DW,
which can be regarded as the edge counterpart of the AHE in bulk magnetic WSMs.
Thus we expect that the DW motion in WSMs is sensitive to the application of an external electric field,
which can be tuned via the chemical potential.

While we have taken the simple collinear DW texture in our analysis,
there can be rich types of DW textures depending on magnetic anisotropy,
such as spiral, head-to-head, vortex DWs, etc.
It is suggested in terms of effective field theory
that WSMs intrinsically host anisotropic spin correlation due to spin-momentum locking \cite{Araki_spincorrelation},
which can lead to significant difference in DW excitation energies and bound-state properties
among those DW textures \cite{Yoshida}.

\textit{Note added.} ---
While preparing this manuscript,
we became aware of the recent related work treating the effect of the axial magnetic field
in the interface of WSMs \cite{Grushin}.
Their numerical calculation results on the LLs, Fermi arc, and the equilibrium current are consistent with
our analytical expressions.

\

\acknowledgments{
The authors thank O.~A.~Tretiakov for helpful discussions.
Y.~A. is supported by JSPS KAKENHI Grant Number JP15H06023.
K.~N. is supported by JSPS KAKENHI Grant Numbers JP26400308 and JP15H05854.}

\appendix

\section{Detailed derivation of bound states}
Here we give a detailed explanation about the derivation of spectrum and wavefunctions
of the bound states around the DW by solving Eq.~(\ref{eq:Weyl-DW}),
which explicitly reads
\begin{align}
(k_z +\epsilon)v(x) &= i\left[-\partial_x +k_y -k_\Delta \tanh\frac{x}{L} \right]u(x) \label{eq:Dirac-vu} \\
(k_z -\epsilon)u(x) &= i\left[\partial_x +k_y -k_\Delta \tanh\frac{x}{L} \right]v(x) . \label{eq:Dirac-uv}
\end{align}

\subsection{The ``Fermi arc'' modes}
Equations (\ref{eq:Dirac-vu}) and (\ref{eq:Dirac-uv}) can be easily solved if the dispersion is linear in $k_z$.
If $\epsilon = k_z$, Eq.(\ref{eq:Dirac-uv}) reduces to the first-order linear differential equation for $v(x)$,
\begin{align}
\left[\partial_x +k_y -k_\Delta \tanh\frac{x}{L} \right]v(x) =0, \label{eq:Dirac-v}
\end{align}
which yields the solution
\begin{align}
v(x) = v_0 \exp\left[-k_y x +k_\Delta L \ln\cosh\frac{x}{L}\right] \label{eq:solution-v}
\end{align}
with $v_0$ a constant.
Since this solution shows the asymptotic behavior
\begin{align}
v(x) \sim v_0 e^{(-k_y\pm k_\Delta)x}
\end{align}
and exponentially diverges in the limit $x\rightarrow \pm\infty$,
the constant $v_0$ should be zero for the normalizability (here $\Delta_0$ is limited to a positive value).
Then we obtain another equation for $u(x)$ from Eq.~(\ref{eq:Dirac-vu}),
\begin{align}
\left[-\partial_x +k_y -k_\Delta \tanh\frac{x}{L}\right]u(x) =0. \label{eq:Dirac-u}
\end{align}
This equation gives the solution
\begin{align}
u(x) = u_0 \exp\left[k_y x -k_\Delta L \ln\cosh\frac{x}{L}\right], \label{eq:solution-u}
\end{align}
which exponentially converges to zero in the limit $x\rightarrow \pm \infty$ as long as $|k_y|<k_\Delta$.

If $\epsilon= -k_z$, on the other hand, Eq.~(\ref{eq:Dirac-vu}) reduces to Eq.~(\ref{eq:Dirac-u}),
and yields the solution given by Eq.~(\ref{eq:solution-u}).
Then Eq.~(\ref{eq:Dirac-uv}) leads to a nonlinear equation for $v(x)$,
\begin{align}
\left[\partial_x +k_y -k_\Delta \tanh\frac{x}{L} \right]v(x) = -2i k_z u_0 e^{k_y x -k_\Delta L \ln\cosh\frac{x}{L}}.
\end{align}
Taking the ansatz
\begin{align}
v(x) = u_0 f(x) e^{-k_y x +k_\Delta L \ln\cosh\frac{x}{L}}, \label{eq:solution-v2}
\end{align}
we find
\begin{align}
f'(x) = -2i k_z \exp\left[2k_y x -2k_\Delta L \ln\cosh\frac{x}{L}\right].
\end{align}
Since $f'(x)$ does not change its sign and converges to zero in the limit $x\rightarrow\pm\infty$,
$f(x)$ is a monotonic function in $x$ and converges to constant values $f_\pm$.
Thus we find that $v(x)$ given by Eq.~(\ref{eq:solution-v2}) exponentially diverges in the limit $x\rightarrow\pm\infty$,
hence the dispersion $\epsilon = -k_z$ is forbidden.

Therefore, we find a linearly dispersed mode $E_0(\bfk_\perp) = v_F k_z$ for $k_y \in (-k_\Delta ,k_\Delta)$,
with the wavefunction
\begin{align}
\psi_{0,\bfk_\perp}^{(+)}(x) = A(k_y) \exp\left[k_y x -k_\Delta L \ln\cosh\frac{x}{L}\right] \bmc{1}{0},
\end{align}
where $A(k_y)$ is the normalization constant that shall be defined below.
For the opposite chirality, the solution can be obtained in the similar way,
with the linear dispersion $E=v_F k_z$ and the wavefunction
\begin{align}
\psi_{0,\bfk_\perp}^{(-)}(x) = A(k_y) \exp\left[-k_y x -k_\Delta L \ln\cosh\frac{x}{L}\right] \bmc{0}{1}.
\end{align}

Here we can determine the normalization constant $A(k_y)$ by
\begin{align}
1 &\overset{!}{=} \int_{-\infty}^{\infty} dx \ |\psi_{0,\bfk_\perp}^{(\tau_z)}(x)|^2 \\
 &= A^2(k_y) \int_{-\infty}^{\infty} dx \ e^{2 \tau_z k_y x} \exp\left[-2 k_\Delta L \ln\cosh\frac{x}{L}\right] \\
 &= \frac{1}{2} \sum_{\pm} \frac{A^2(k_y)}{k_\Delta \pm k_y} F\left(1,2 k_\Delta L;\; 1+k_\Delta L \pm k_y L;\frac{1}{2}\right). \label{eq:normalization}
\end{align}
$F(a,b;c;z)$, explicitly denoted by $_2 F_1(a,b;c;z)$, is the ``hypergeometric function'' defined by
\begin{align}
F(a,b;c;z) \equiv \sum^{\infty}_{n=0} \frac{(a)_n (b)_n}{(c)_n} {z^n}{n!}
\end{align}
with
\begin{align}
(a)_n = a (a+1) \cdots (a+n-1).
\end{align}
We should note that $A(k_y)$ has the dimension of $[L^{-1/2}]$.

\subsection{The other bound states}
In order to find the bound states other than the ``Fermi arc'' modes shown above,
we need to solve the second-order differential equation Eq.~(\ref{eq:Weyl-Schrodinger}),
namely
\begin{align}
\left[ \partial_x^2 + \frac{k_\Delta}{L}\sech^2\frac{x}{L} - \left(k_y -k_\Delta \tanh\frac{x}{L}\right)^2 +\epsilon^2 -k_z^2 \right] u(x) =0. \label{eq:Weyl-2}
\end{align}
As mentioned in the main part of this paper,
$u(x)$ shows the asymptotic behavior
\begin{align}
u(x) \sim e^{\mp \kappa^{\pm} x}
\end{align}
in the limits $x \rightarrow \pm \infty$.
Thus we take the ansatz
\begin{align}
u(x) \equiv \left(1+e^{\frac{2x}{L}}\right)^{\frac{1}{2}\kappa^+ L} \left(1+e^{-\frac{2x}{L}}\right)^{\frac{1}{2}\kappa^- L} \tilde{u}(x),
\end{align}
with the boundary condition $\tilde{u}(x\rightarrow\pm\infty) \sim O(1)$.
Using the new variable $\zeta \equiv \frac{1}{2}[1-\tanh(x/L)]$,
Eq.~(\ref{eq:Weyl-2}) reduces to the hypergeometric differential equation,
\begin{align}
\left[ \zeta(1-\zeta)\partial_\zeta^2 + \left( c-(1+a+b)\zeta \right)\partial_\zeta -ab \right] \tilde{u}(\zeta) =0,
\end{align}
where
\begin{align}
a &= (\bar{\kappa} +k_\Delta)L +1 \\
b &= (\bar{\kappa} - k_\Delta)L \\
c &= \kappa^+ L +1,
\end{align}
with $\bar{\kappa} = (\kappa^+ + \kappa^-)/2$.
This equation has two linearly independent solutions:
\begin{align}
\tilde{u}^{\mathrm{I}}(\zeta) &= F(a,b;c;\zeta) \\
\tilde{u}^{\mathrm{II}}(\zeta) &= \zeta^{1-c} F(1+a-c,1+b-c;2-c;\zeta).
\end{align}
Since $\tilde{u}^{\mathrm{II}}$ leads to a singularity in $u$ at $\zeta=0 \ (x=\infty)$,
the solution is given solely by $\tilde{u}^{\mathrm{I}}$.

The parameters $a,b$ and $c$ are limited by the boundary condition at $\zeta=1 \; (x=-\infty)$.
$\tilde{u}^{\mathrm{I}}(\zeta)$ can be rewritten as
\begin{align}
\tilde{u}^{\mathrm{I}}(\zeta) = (1-\zeta)^{c-a-b} F(c-a,c-b;c;\zeta)
\end{align}
by Euler's transformation.
Since $c-a-b = -\kappa^- L$ is negative, the factor $(1-\zeta)^{c-a-b}$ diverges at $\zeta=1$.
Therefore, to meet the boundary condition at $\zeta=1$,
the equation $F(c-a,c-b;c;1) =0$ should be satisfied.
From Gauss's theorem, this equation reads
\begin{align}
& F(c-a,c-b;c;1) = \frac{\Gamma(c)\Gamma(a+b-c)}{\Gamma(a)\Gamma(b)} \\
 &\quad \quad = \frac{\Gamma(\kappa^+ L +1) \Gamma(\kappa^- L)}{\Gamma((\bar{\kappa}+k_\Delta)L +1) \Gamma((\bar{\kappa}-k_\Delta)L)} =0. \label{eq:F0}
\end{align}
Since the gamma function $\Gamma(z)$ is nonzero, and diverges if $z$ is a non-positive integer,
Eq.~(\ref{eq:F0}) yields the quantization condition
\begin{align}
(\bar{\kappa}-k_\Delta)L = -N. \quad (N=0,1,2,\cdots)
\end{align}
Substituting $\kappa_N^\pm = \left[ k_z^2 + (k_y \mp k_\Delta)^2 - \epsilon_N^2 \right]^{1/2}$
and solving the relation for $\epsilon_N$,
we obtain the eigenvalue
\begin{align}
\epsilon_N(\bfk_\perp) = \pm \sqrt{k_z^2 + \left(\frac{k_\Delta^2}{\bar{\kappa}_N^2} -1 \right)(\bar{\kappa}_N^2 - k_y^2)} , \label{eq:energy}
\end{align}
with the eigenfunction
\begin{align}
u_N(\zeta) &= \zeta^{\frac{1}{2}\kappa_N^+ L} (1-\zeta)^{\frac{1}{2}\kappa_N^- L} \\
 & \quad \quad \times F(-N,-N+2k_\Delta L +1; \; 1+\kappa_N^+ L; \zeta). \nonumber
\end{align}
Using Eq.~(\ref{eq:Dirac-vu}),
the lower component is straightforwardly obtained as
\begin{align}
v_N(\zeta) &= i \frac{\kappa_N^+ -k_\Delta +k_y}{k_z + \epsilon_N} \zeta^{\frac{1}{2}\kappa_N^+ L} (1-\zeta)^{\frac{1}{2}\kappa_N^- L} \\
 & \quad \quad \times F(-N+1,-N+2k_\Delta L ; \; 1+\kappa_N^+ L; \zeta). \nonumber
\end{align}
The normalization constants can be obtained accordingly,
which we will not calculate in detail
because it requires integration over the product of hypergeometric functions, which cannot be evaluated analytically.

Using the equations
\begin{align}
\frac{1}{2}(\kappa_N^+ + \kappa_N^-) &= \bar{\kappa}_N = k_\Delta - \frac{N}{L} \label{eq:app-k1} \\
(\kappa_N^+)^2 - (\kappa_N^-)^2 &= -4 k_\Delta k_y,
\end{align}
the decay rates $\kappa_N^\pm$ are given by
\begin{align}
\kappa_N^\pm &= \bar{\kappa}_N \pm \frac{k_\Delta k_y}{\bar{\kappa}_N}.
\end{align}
Since we are interested in the bound state solutions,
we require $\kappa_N^\pm$ to be real and positive.
Thus we obtain the condition for $k_y$,
\begin{align}
|k_y| < \frac{\bar{\kappa}_N^2}{k_\Delta}, \label{eq:app-k-cond}
\end{align}
otherwise the solution gives the extended state.

In summary,
the wavefunction for the chirality $\tau_z=+1$ is given by
\begin{align}
\psi_{N,\bfk_\perp}^{(+)}(x) &= (1+e^{\frac{2x}{L}})^{\frac{1}{2}\kappa^+_N L} (1+e^{-\frac{2x}{L}})^{\frac{1}{2}\kappa^-_N L} \tilde{\psi}_{N,\bfk_\perp}^{(+)}(x) \\
\tilde{\psi}_{N,\bfk_\perp}^{(+)}(x) &= \bmc{F\left(-N,-N+2k_\Delta L +1; \ 1+ \kappa^+_N ; \ \zeta \right)}{i \eta_N F\left(-N+1,-N+2k_\Delta L ; \ 1+ \kappa^+_N ; \ \zeta \right)}, \nonumber
\end{align}
where the coefficient $\eta_N$ is defined by
\begin{align}
\eta_N = \frac{\kappa^+_N -k_\Delta +k_y}{k_z+\epsilon_N}.
\end{align}


\

\end{document}